\documentstyle[aps,prl,psfig,array]{revtex}

\newcommand{\be}{\begin{equation}}
\newcommand{\ee}{\end{equation}}

\newcommand{\bex}{\begin{eqnarray}}
\newcommand{\eex}{\end{eqnarray}}
\begin{document}

\title{Securing quantum bit commitment through reverse quantum communication}
\author{R. Srikanth\thanks{e-mail: srik@iiap.ernet.in}}
\address{Indian Institute of Astrophysics, 
Koramangala, Bangalore- 34, Karnataka, India.}
\maketitle \date{}


\begin{abstract}
Einstein-Podolsky-Rosen- (EPR) and the more powerful Mayers-Lo-Chau attack 
impose a serious constraint on quantum bit commitment (QBC). As a way to 
circumvent them, it is proposed that the quantum system encoding the commitment 
chosen by the committer (Alice) should be initially prepared in a seperable 
quantum state known to and furnished by the 
acceptor (Bob), rather than Alice. Classical communication is used to conclude 
the commitment phase and bind Alice's subsequent unveiling. 
Such a class of secure protocols can be built upon currently proposed QBC 
schemes impervious to a simple EPR attack. A specific scheme based on the
Brassard-Cr\'epeau-Josza-Langlois protocol is presented here as an example. 
\end{abstract}

\section{Introduction}
Quantum cryptography is concerned with harnessing the
principles of quantum mechanics to create secure
cryptosystems \cite{wie83,bb84}. It can broadly be
divided into quantum key distribution (QKD) \cite{bb84,eke91}, which is 
concerned with secure sharing of cryptographic keys, 
and a host of other schemes,
such as quantum coin tossing \cite{brcr91,bcjl,ard95},
quantum oblivious mutual identification \cite{crsl95}, quantum 
oblivious transfer \cite{cre94} and ``two party secure computation" 
(TPSC) \cite{kil88}, essentially concerned with secure processing
of private information to reach a public decision. The latter schemes depend
on the validity of quantum bit commitment (QBC) \cite{bb84,bcjl}, a 
quantum cryptographic
primitive for secure information processing. In a concrete realization
of bit commitment, Alice writes 0 or 1 on a note, puts it
in a safe, which she hands over to Bob. Upon Bob choosing to enter the
transaction, she gives him the key to the safe. The main point is that
Alice cannot cheat by changing her mind after handing Bob the safe, nor 
can Bob cheat by finding
out about Alice's decision unless she gives him the key. A secure bit commitment
is one which is (at least, exponentially) binding on Alice {\em and} 
unconditionally concealing (of her commitment) from Bob and thus prevents
either party from cheating. 

That entanglement can undermine QBC was first realized by Bennett and Brassard
\cite{bb84}, who pointed out that the BB84 scheme \cite{bb84} was insecure
against an Einstein-Podolsky-Rosen (EPR) attack by Alice \cite{epr}, i.e., a
deception where she sends Bob part of
entangled photons instead of ones in a definite polarization
state, and waits until after the initial phase of the protocol to
measure the part she retains. A subsequent proposition, namely the BCJL 
protocol \cite{brcr91,bcjl}, though impervious to an EPR attack, 
is nevertheless rendered insecure by an entanglement-based attack
independently uncovered by Lo and Chau \cite{loc97,loc98} and 
Mayers \cite{may97}. 
The essence of their proof is that if the protocol is
secure against Bob, Alice can cheat by supplying him a pure state entangled
system, and switching her commitment by local unitary operations. 
It has come to be accepted that QBC simultaneously secure against
both Alice and Bob is impossible, though a trade-off is permitted \cite{spe01}. 

Almost all QBC schemes we know conform to a model wherein the quantum
system encoding the committer's (Alice's) commitment $b$ is prepared by
her. She is the sender. The party that accepts her commitment, the ``acceptor"
called Bob, is the receiver of the quantum information. 
The present article explores whether QBC can be made secure if Bob,
rather than Alice, prepares and sends the initial state of the encoding
quantum system, and Alice is the receiver who ``inscribes"
her commitment on the state furnished by Bob. The motiviation to do so is quite
straightforward, given that the insecurity of QBC so far stems from Alice 
preparing the initial state as she likes.

\section{The modified BCJL scheme}
In a typical QBC scheme, Alice prepares a quantum state consisting of
a pre-agreed number $n$ of photons in
a pure product state determined by her commitment $b$ ($\in \{0, 1\}$). Four
possible preparations of a photon
are permitted: in the rectilinear basis (denoted +),
with horizontal (denoted 0) or vertical (denoted 1) 
polarization; else, in the diagonal basis \mbox{(denoted
$\times$)} with polarization oriented at $45^{\circ}$ (denoted 0) or 
$135^{\circ}$ (denoted 1). 
She sends these photons to Bob as a piece of evidence of her commitment.
If Bob signals his acceptance to enter the transaction, she unveils $b$ and
the preparation information of her photons, which Bob then verifies by
measuring their polarizations. 
In order that Bob should not cheat, the density matrices of the system given
to Bob corresponding to $b = 0, 1$
\mbox{-- namely $\rho^B_0, \rho^B_1$--} should be 
(almost) indistinguishable. On the other hand, the preparation of the
photons she sends Bob should be binding on her. As noted earlier, if a 
Mayers-Lo-Chau (MLC) attack
can be launched, then these two requirements are mutually exclusive 
\cite{loc97,loc98,may97}. 
  
A simple way to circumvent an EPR or MLC attack is to prevent Alice from
launching one. The main point of the present article is that
Bob can achieve this by preparing a
seperable $n$-particle quantum state and sending it to Alice to inscribe
her commitment. We need to show that any subsequent action by Alice will
not permit her to launch an attack. Then, a 
classical communication from Alice is sufficient
to guard against her cheating. The hallmarks of the class of
protocols we envisage are (1) a reverse quantum communication, wherein Alice 
and Bob have exchanged their traditional
 sender-receiver roles, and (2) a classical 
communication from Alice to signal the end of the commitment phase. 

We present a version of the BCJL protocol modified to include these two
features. The proof of its security is given thereafter.  
The modified BCJL scheme
consists of two phases, the commitment phase and unveiling 
phase, enumerated below. An intervening phase of arbitrary duration, referred
to as the holding phase, is implicit but ignored in the analysis.
Since we only wish to present a proof of principle, 
discussion on error correction is not included here.
\begin{enumerate} 
\item {\bf Commitment phase}:
\begin{enumerate}
\item Alice and Bob agree upon an $n$-bit code ${\cal C}$ 
(with some required properties).
\item They also agree upon a random $n$-bit string $r \in \{0, 1\}^n$.
\item Bob chooses a random $n$-bit string $R_B \in {\cal C}$. 
\item He chooses a random $n$-bit string $\eta \in 
\{+, \times\}^n$ and prepares the state $|R_B\rangle_{\eta} = 
|R_B(1)\rangle_{\eta(1)}\otimes\cdots\otimes|R_B(n)\rangle_{\eta(n)}$ and
sends it to Alice.  
\item She chooses a random $n$ bit string $\theta \in 
\{+, \times\}^n$ and measures Bob's photons in bases $\theta$, obtaining
outcomes $R_A \in \{0, 1\}^n$. 
\item To encode her commitment $b$, Alice checks whether 
$\{R_A \in {\cal C}~|~r \odot R_A = b\}$, where
the symbol $\odot$ denotes scalar product modulo 2, or the parity of the
bitwise AND operation. If the check succeeds (fails), she excludes a photon
at a randomly chosen position $x$ where she obtained outcome 0 (1), to obtain
$R_A^{\prime}$ such that \mbox{$\{R_A^{\prime} \in {\cal C}~|~r \odot 
R^{\prime}_A = b\}$}.
\item She communicates to Bob the value of $x$.
This announcement serves as a piece of evidence of her commitment. 
\end{enumerate}
\item {\bf Unveiling phase}:
\begin{enumerate}
\item Alice announces $b$, $R_B$ and $\theta$. 
\item Bob verifies that:
$r \odot R_B^{\prime} = b$.
\item He also verifies that whenever $\eta(i) = \theta(i)$,
$R_A(i) = R_B(i)$.
\end{enumerate}  
\end{enumerate}  

Even though Bob prepares and thus knows the state $|R_B\rangle_{\eta}$
he sent Alice, without access to her outcomes where $\eta(i) \ne \theta(i)$,
he cannot deduce $b$. In fact, he doesn't even know where $\eta$ and $\theta$
don't match. Further, Alice need not fear that by sending
in an entangled state rather than a seperable state 
$|R_B\rangle_{\eta}$, Bob can hope to get
information about her measurement outcomes. The very nature of quantum
measurement (assumed to be a von Neumann projection or, more generally,
a positive operator valued measure)
prohibits Bob from knowing anything about her action, because
Bob's local density matrix is unaffected by Alice's
measurement. Therefore, even empowered with a quantum computer, Bob cannot hope
to deduce her outcomes by observations local
to him. One way to view this is that if this were not the case, then Alice
could transmit superluminal signals to Bob according to her choice of
$\theta$ \cite{ghi88}. Another deterrent for Bob, as shown below, is that 
Alice could launch an MLC attack 
against him if the system remains entangled with a
hidden system on his side at the time
of her measurement. Therefore the modified protocol is indeed secure against
Bob. 

Since Bob knows the exact pure, seperable
state he sent her, any entanglement-based attack
cannot be launched by Alice. Moreover the no-cloning theorem \cite{woo82}
prevents her from knowing the exact state Bob sent her. Were this not so, she
could find out the exact state Bob sent her and
cheat by unveiling some $\theta(i)$ in the wrong basis and claim any
outcome she likes. Thus, if she decides to
switch her commitment in the unveiling phase, the best she can do is to 
flip some $R^{\prime}_B(i)$ with outcome 1, and
announce the dishonest $R^{\prime}_B$ to Bob.  But her
probability for getting away without Bob noticing is 1/2. Therefore, if the
scheme is implemented on $s$ rounds, her probability for cheating successfully
falls exponentially as $\left(\frac{1}{2}\right)^{s}$. 
Alice's classical communication in step 1(g) serves the crucial 
dual role of signaling Bob that the commitment phase is over,
as well as binding Alice's to-be-unveiled commitment. 
This step is necessary because he cannot deduce based on local measurements
whether she has measured or not. On the other, a mere intimation on her part
is clearly insufficient, since she could simply lie that she measured,
while actually intending to delay the decision on her commitment. 
It must bind her future 
unveiling, but without betraying $b$. Therefore, we expect that any proposed 
scheme that constrains outcomes of measurements-- with the result that
$\rho^B_0$ and $\rho^B_1$ slightly differ (eg., the BCJL scheme)-- rather than 
measurement basis (eg., the BB84 scheme),
permits a secure version along the lines given above. 

\section{Entanglement weakens security against Alice}
If Bob were to send Alice half of
EPR pairs instead of the pure state $|R_B\rangle_{\eta}$, 
the above protocol would still be secure against EPR attacks by
Alice. However, Alice can launch an MLC attack.
To this end, she does not execute the measurement in 1(e).
For step 1(g), Alice chooses a random photon for exclusion.
Suppose each of Alice's remaining photons is half of an EPR pair in the state
$|\phi\rangle = \frac{1}{\sqrt{2}}(|0\rangle_+|0\rangle_+ + 
|1_+\rangle|1\rangle_+)$, the other half being sent to Bob. 
The $2(n-1)$ particle state is given by
\be
\label{2n}
|\Phi\rangle_{AB} = 2^{-(n-1)/2}
\sum_{j=1}^{2^{n-1}}|j\rangle_A\otimes|j\rangle_B
\ee 
in the $+$ basis. The register $A$ is sent to Alice, while $B$ remains with
Bob.  Since the BCJL permits $N \equiv 
2^{n-1}\cdot2^{n-2}$ states encoding a given commitment, she augments register
$A$ she received from Bob by adding ancilla $C$, so that ${\cal H}_C \otimes
{\cal H}_A$, the Hilbert space of the combined $CA$
system, has the dimension $N$ \cite{hug93}. 
\be
\label{3n}
|\Psi\rangle_{CAB} = 2^{-(n-1)/2}\sum_{j=1}^{2^{n-1}}|e_j\rangle_C\otimes 
|j\rangle_A\otimes|j\rangle_B 
\equiv 2^{-(n-1)/2}\sum_{j=1}^{N}|f_j\rangle_{CA}\otimes|j\rangle_B,
\ee 
where, for $j>2^{n-1}$, we set $|j\rangle_B = 0$, or null state, but retain the
the $|j\rangle_B$'s as before for $j \le 2^{n-1}$. 

We denote an ensemble of states in the BCJL scheme encoding
commitment $b=0$ by 
$\{\sqrt{p_j}|0_j\rangle\}$ where $\sum_jp_j|0_j\rangle\langle 0_j| \equiv
\rho^B_0 \approx {\rm Tr}_A\left(|\Phi\rangle_{AB}\langle\Phi|_{AB}\right)$. 
To produce it, she determines the $N \times N$ 
unitary mixing matrix $U^0$ such that:
\be
\label{mix}
2^{-(n-1)/2}|j\rangle_B = \sum_{k=1}^N U^0_{jk}\sqrt{p_j}|0_k\rangle_B.
\ee 
Denoting
$|g_j\rangle_{CA} \equiv \sum_{k=1}^NU^0_{jk}|f_k\rangle_{CA}$, 
we rewrite Eq. (\ref{3n}) as:
\be
\label{cheat}
|\Psi\rangle_{CAB} = \sum_{j=1}^{N}\sqrt{p_j}
|g_j\rangle_{CA}\otimes|0_j\rangle_B.
\ee
By measuring in the $\{|g_j\rangle\}$ basis of the $AC$ system, she produces a 
state compatible with $b=0$. On the other hand, 
to convince Bob she is committed to $b=1$,
starting from Eq. (\ref{cheat}), Alice 
applies a unitary transformation to the $CA$ register to produce an ensemble
$\{\sqrt{q_j}|1_j\rangle_B\}$ where $\sum_jq_j|1_j\rangle\langle 1_j| \equiv
\rho^B_1 \approx {\rm Tr}_A\left(|\Phi\rangle_{AB}\langle\Phi|_{AB}\right)$ 
on Bob's side. Such a transformation exists since $\rho^B_0 \approx \rho^B_1$ 
\cite{loc97,loc98,may97}.
Alternatively, she can directly determine the $N \times N$ unitary
mixing matrix $U^1$ of her incremented system needed to generate
$\{\sqrt{q_j}|1_j\rangle_B\}$ state. 

The attack works if Alice knows the exact entangling state ($|\phi\rangle$,
in this case) Bob used. 
Nevertheless, it is interesting that the danger of an MLC attack exists even 
when Bob prepares the state that is to encode
Alice's commitment. The main lesson is that
the use of entanglement in any form undermines the security of QBC against
Alice. 

\section{Conclusion}

The vicissitudes of QBC's fate are a bit remarkable and indicate its subtle
nature. Are there simple fundamental causes why the present version of
QBC succeeds? In answer, one might say, as in QKD: Heisenberg uncertainty, 
quantum no-cloning and ``causality"-- that Bob cannot deduce Alice's action
without classical input from her.
It is interesting that essentially the
same properties of quantum information that guarantee security
between collaborating parties in quantum key distribution \cite{bb84} do the
same even when the two parties are mutually distrustful.

\end{document}